\documentclass[dvipdfmx]{apex}
\usepackage{txfonts}
\usepackage{braket}
\usepackage{bm}

\def\HC{{H_\mathrm{c}}}
\title{Theoretical evaluation on the temperature dependence of magnetic anisotropy constants of $\rm \textbf {Nd}_\textbf{2} \textbf {Fe}_\textbf {14} \textbf {B} $: Effects of exchange field and crystal field strength\ --}
\author{Ryo Sasaki, Daisuke Miura, and Akimasa Sakuma}
\inst{Department of Applied Physics, Tohoku University, Aoba-ku, Sendai 980-8579, Japan}

\makeatletter
\AtBeginDocument{\@ifpackageloaded{natbib}{\ifNAT@numbers\if@filesw\immediate\write\@auxout{\string\global\string\NAT@numberstrue}\fi\fi}{}}
\makeatother
\begin{document}

%\abst{This document briefly provides instructions on how to prepare your manuscript in \LaTeX\ format. As regards general instructions for preparing manuscripts, please refer to ``Instructions for Preparation of Manuscript", which is available at our Web site (http://apex.ipap.jp/).}
\abst{To identify the possible mechanism of coercivity ($\HC$) degradation of Nd-Fe-B sintered magnets, we study the roles of the exchange field acting on the 4f electrons in Nd ions and  theoretically investigate how the variation of the exchange field affects the values of the magnetic anisotropy constants $K_1$ and $K_2$.  We find that, with decreasing exchange field strength,  both values decrease as a result of the lower asphericity of the 4f electron cloud,  indicating  that the local anisotropy constants might become small  around the grain boundaries where the exchange fields are decreased owing to the smaller coordination number.}

\maketitle

%You can use this file as a template to prepare your manuscript for \textit{Applied Physics Express} (APEX)\cite{apex}. No sections or appendices should be given. Key words are not necessary. Use the \verb|twocolumn| option to estimate the length of your manuscript because APEX has  a limitation of three printed pages. If the \verb|txfonts| package is available in your \LaTeX\ system, you can estimate the length more accurately. However, prepare a one-column version when you submit your manuscript.

%Copy the \verb|apex.cls|, \verb|cite.sty|, and \verb|overcite.sty| onto an arbitrary directory under the texmf tree, for example, \verb|$texmf/tex/latex/ipap|. If you have already obtained the \verb|cite.sty| and \verb|overcite.sty|, you do not need to copy them.

%Many useful commands for equations are available because the \verb|apex.cls| automatically loads the \verb|amsmath| package. Please refer to reference books on \LaTeX\ for details on the \verb|amsmath| package.

Nd-Fe-B sintered magnets\cite{1,2,3} have the largest maximum energy product among the current magnets and have been widely used for magnetic devices such as voice coil motors in magnetic recording systems. Recently, because of the rapidly growing interest in electric vehicles, much effort has been made to suppress the degradation of the coercivity ($\HC$) of  Nd-Fe-B magnets.  However, from an industrial viewpoint, reduction in the usage of Dy is strongly desired, because Dy is a rare metal and the magnetization of the Nd-Fe-B magnets decreases by  substituting Dy with Nd owing to the antiparallel coupling between Dy and Fe moments.  Realizing Dy-free high-performance Nd-Fe-B magnets requires  a further increase of $\HC$ in the Nd-Fe-B system  by microstructure optimization,\cite{4,5,6,7,8,9,10,11} and therefore, establishing the microscopic foundation for the coercivity mechanism is desired. From a theoretical viewpoint, many works\cite{12,13,14,15,16} have  focused on the change of magnetic anisotropy constants around the grain boundary surfaces as a result of the stresses, defects, and change of spatial symmetry. In addition,  micromagnetic model calculations have shown that the surface $c$-plane anisotropy can  drastically decrease the coercivity.\cite{17}  For these reasons, evaluation of the local anisotropy constants around the grain boundaries and determining their temperature dependence are important to investigate the degradation of the coercivity.\par

With regard to the magnetic anisotropy of rare earth (RE)  transition metal  compounds, it is believed that the 4f electrons in RE ions are responsible for the main part of the magnetic anisotropy and that the crystalline electric field (CEF) acting on the 4f electrons dominates this property.\cite{18}  Under the assumption that the exchange field on the 4f electrons is strong enough, by using the CEF parameter $A_2^0$, the leading anisotropy constant $K_1$ can be approximately described by $K_1=-3J(J-1) \alpha \langle {\hat r}^2 \rangle A_2^0 N_{\mathrm R}$, where $\alpha$ is the Stevens factor, $J$ is the total angular momentum, $\langle {\hat r}^2 \rangle$ is the average of ${\hat r}^2$ over the radial wave function of the 4f electrons, and $N_{\mathrm R}$ is the density of RE ions. Note that the CEF parameter is easily affected by circumstances especially  around the grain boundaries.  Actually, Moriya et al.\cite{15} predicted by using a first-principles calculation that the CEF parameter $A_2^0$ exhibits a negative value at the (001) surface of $\rm Nd_2 Fe_{14} B$ .  According to Ref. 17, this may lead to a drastic degradation of $\HC$.\par

Worth mentioning here is that the exchange field acting on the 4f electrons from surrounding Fe spins  can also be changed as a result of the situation surrounding the RE ion.  It is natural to consider that the exchange fields are weak  around the grain boundaries compared to those in the bulk.  In this case, the above expression for $K_1$ no longer applies because  the precondition that the exchange field is strong enough no longer holds.  Especially, the decrease of the exchange field possibly has a significant influence on the temperature dependence of the anisotropy constants because of thermal fluctuations of the 4f moments.  Based on this viewpoint, we focus on the roles of the exchange field on the 4f-related anisotropy constants, and we  theoretically investigate how the magnetic anisotropy is affected by a variation of exchange fields and the CEF acting on the 4f electrons.  To this aim, the temperature-dependent anisotropy constants $K_1 (T)$ and $K_2 (T)$ are calculated based on crystal field theory. \par

The calculation method we use here is based on conventional crystal field theory for the 4f electronic system in the RE ions. The total Hamiltonian for 4f electrons is given as
$\hat{\mathcal{H}} = \hat{\mathcal{H}}_\mathrm{ion} + \hat{\mathcal{H}}_\mathrm{CEF} + \hat{\mathcal{H}}_\mathrm{mag}$,
where $\hat{\mathcal{H}}_\mathrm{ion}$ describes the intra-atomic interactions in an RE ion,
$\hat{\mathcal{H}}_\mathrm{CEF}$ denotes the CEF part representing the electrostatic field acting on the 4f electrons from the surrounding charge,
and $\hat{\mathcal{H}}_\mathrm{mag}$ denotes the effective exchange interactions between the 4f and Fe moments.
When Nd ions are involved, only the ground $J$ multiplet can be considered as the 4f states because 4f electrons of the Nd ion have large spin-orbit coupling compared to the CEF splitting.  In this case,
$\hat{\mathcal{H}}_\mathrm{CEF}$ can be written as $\hat{\mathcal{H}}_\mathrm{CEF} = \sum_{l,m} B_l^m \hat{\mathcal{O}}_l^m$,
where $B_l^m=\theta_l \langle r^l \rangle A_l^m$ is the CEF coefficient and $\hat{\mathcal{O}}_l^m$ is the Stevens operator expressed by the multinomial of the orbital angular momentum operators.
$\theta_l$ is the Stevens factor $\alpha,$ $\beta,$ or $\gamma$ for $l=2$, 4, or 6, respectively, and $A_l^m$ are the CEF parameters.
In this work, we take into account only $B_2^0,$ $B_4^0$, and $B_6^0$, for simplicity.
Within the ground $J$ multiplet, $\hat{\mathcal{H}}_\mathrm{mag}$ can be expressed as $\hat{\mathcal{H}}_\mathrm{mag} = 2(g_J-1) \hat{\bm J} \cdot {\bm H}_\mathrm{ex}$,
where $g_J$ is the Lande factor and $\bm{H}_\mathrm{ex}$ is the effective exchange field reflecting the exchange interaction with surrounding Fe spins.
%In principle, $\bm{H}_\mathrm{ex}$ itself is temperature dependent through a self-consistent equation for the molecular field approximation.
In principle, $\bm{H}_\mathrm{ex}$ depends on temperature through a self-consistent equation for a molecular field approximation in the full system of $\mathrm{Nd_2 Fe_{14} B}$.
However, in the present work,
%we have skipped such the treatment because the CEF is too small to influence on the $\bm{H}_\mathrm{ex}$.
we assume that $\bm{H}_\mathrm{ex}$ is proportional to a molecular field of Fe spins $\braket{\bm S_\mathrm{Fe}}$ and that the molecular field is described by the Heisenberg model for the Fe-spin system.
Solving the self-consistent equation for $\braket{\bm S_\mathrm{Fe}}$ under the condition that the spin of the Fe ion is 1 and that the Curie temperature $T_\mathrm{c}$ is 583 K,
we have confirmed that $\braket{\bm S_\mathrm{Fe}} $ reproduces well the temperature dependence of the magnetization of $\mathrm{Y_2 Fe_{14} B}$.\cite{19}
%In calculating $B_S (T)$, we assumed that the spin of Fe ion is 1 and Curie temperature $T_c$ is 583 [K].
%we have confirmed that $B_{S=1} (T)$ reproduces well the temperature dependence of magnetization $M(T)$ of $\rm Nd_2 Fe_{14} B$.\cite{19}
%Thus we have introduced temperature dependency in ${\mathbf H}_{\rm ex}$ by ${\mathbf H}_{\rm ex} (T) = H_{\rm ex} B_{S=1} (T) {\mathbf e}$
Thus we have introduced temperature dependency in $\bm{H}_\mathrm{ex}$ by $\bm{H}_\mathrm{ex} (T) = H_\mathrm{ex} \braket{\bm S_\mathrm{Fe}}$,
where $H_\mathrm{ex}$ is the strength parameter of the effective exchange field. % and $\bm e$ means a unit vector indicating the direction of $\bm{H}_\mathrm{ex}$.
Here, the direction of total magnetization is assumed to coincide with that of $\bm{H}_\mathrm{ex}$, because the total magnetization is mostly governed by the Fe moments even when the direction of 4f moments differs from that of the Fe moments. \par

By defining $\theta$ as the angle between $\bm{H}_\mathrm{ex}(T)$ (total magnetization) and the $c$ axis of the crystal lattice, the free energy of the 4f electronic system can be written as
\begin{equation}
F(\theta,T) = -k_{\rm B} T \ln {\rm Tr} \exp \left[ -\frac{\hat{\mathcal{H}}(\theta,T)}{k_{\rm B} T} \right] \nonumber,
\end{equation}
where the eigenvalues are calculated by diagonalizing $\hat{\mathcal{H}}(\theta,T)$ within the $J=9/2$ subspace ($10\times10$ matrix), and $F(\theta,T)$ is numerically obtained.\par

The 4f-related anisotropy constants $K_1$ and $K_2$ are coefficients of $\sin ^2 \theta$ and $\sin ^4 \theta$ in $F(\theta,T)$. For small $\theta$, $F(\theta,T)$ can be expanded as
\begin{align*}
F(\theta, T) &= F(0, T) + \frac{ \theta ^2}{2} F^{(2)} (0, T) + \frac{ \theta ^4}{4!} F^{(4)} (0, T) + \frac{ \theta ^6}{6!} F^{(6)} (0, T) + \cdots\\
&\equiv F(0, T) + K_1 (T)  \theta ^2 + \left[ -\frac{1}{3} K_1 (T) + K_2 (T) \right] \theta ^4 + \cdots.
\end{align*}
Therefore, $K_1$ and $K_2$ can be expressed by
\begin{align*}
K_1(T) &= \frac{1}{2} F^{(2)} (0, T),\\
K_2 (T)&=\frac{1}{3} K_1 + \frac{1}{4!} F^{(4)} (0, T),
\end{align*}
where we calculate $F^{(2)} (0, T)$ and $F^{(4)} (0, T)$ from finite differences.
%\begin{align*}
%F^{(2)} (0, T) &= -\frac{245F(0, T)-270F(\theta, T)+27F(2\theta, T)-2F(3\theta, T)}{90\theta ^2},\\
%F^{(4)} (0, T) &= -\frac{28F(0, T)-39F(\theta, T)+12F(2\theta, T)-F(3\theta, T)}{3\theta ^4}.
%\end{align*}
\par

\begin{figure}
\begin{center}
\includegraphics[height=6.0cm]{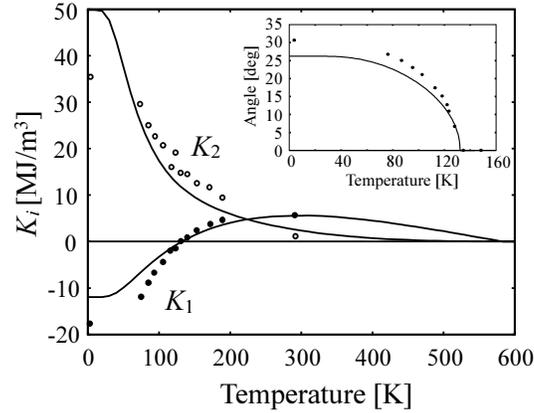}
\end{center}
\caption{Calculated  anisotropy constants $K_1$ and $K_2$ (solid lines) as a function of temperature obtained using $A_2^0=450\ \mathrm{K/a_0{}^2}$, $A_4^0=-45\ \mathrm{K/a_0{}^4}$, $A_6^0=-0.1\ \mathrm{K/a_0{}^6}$,
and $H_\mathrm{ex}=364\ \mathrm{K}$.
The closed and open circles are the measured values \cite{21} of $K_1$ and $K_2$, respectively.
The inset shows the tilting angles of the magnetization from the $c$ axis as a function of temperature together with the experimental results (closed circles). \cite{21}}
\label{f1}
\end{figure}

To perform a quantitative analysis on the magnetic anisotropy constants of $\mathrm{Nd_2 Fe_{14} B}$, we start by reproducing the experimental results of $K_1$ and $K_2$ of bulk $\mathrm{Nd_2 Fe_{14} B}$ by using the present approach.  Figure \ref{f1} shows the calculated results obtained for $K_1 (T)$ and $K_2 (T)$
using
$A_2^0=450\ \mathrm{K/ a_0{}^2}$,
$A_4^0=-45\ \mathrm{K/a_0{}^4}$,
$A_6^0=-0.1\ \mathrm{K/a_0{}^6}$,
and $H_\mathrm{ex}=364\ \mathrm{K}$,\cite{20}
together with the experimental data. \cite{21}
Here, $\mathrm{a_0}$ is the Bohr radius.
In calculating $K_1 (T)$,
we add the Fe sublattice contribution $K_1^\mathrm{Fe} (T)$
deduced from experimental data for $\mathrm{Y_2 Fe_{14} B}$.\cite{19}
This has an almost constant value of $\sim$1 MJ/$\mathrm{m^3}$ below 300 K and decreases to zero when the temperature approaches  $T_\mathrm{c}$.
The difference between two Nd sublattices is not taken into account here.
The inset in Fig. \ref{f1} shows the tilting angle of the magnetization vector calculated from $\frac{\partial F(\theta, T)}{\partial \theta} = 0$. %$\partial F(\theta, T) / \partial \theta = 0$.
Including this spin reorientation behavior,
one can see  agreement between calculated data and experimental ones to a certain degree. \par

It should be noted here that the experimental values of $K_1 (T)$ and $K_2 (T)$ are not obtained by the direct measurement but are usually deduced from magnetization curves or torque curves under finite applied fields.  Furthermore, in our numerical analysis, we made some approximations to express $K_1 (T)$ and $K_2 (T)$.  In this sense, we believe that the CEF and $H_\mathrm{ex}$ parameters reported by Yamada et al. are more reliable, because in their analysis the Zeeman term is considered in addition to the Hamiltonian to ascertain the magnetization curves and to make a direct comparison with the measured ones.
Reflecting this difference, the CEF and $H_\mathrm{ex}$ parameters obtained by Yamada et al. are considerably different from ours.  To examine the availability of the present method and reliability of the parameters obtained above, we also calculated the $K_1 (T)$ and $K_2 (T)$ curves using the parameters given by Yamada et al.\cite{18}
Except for the behavior of $K_2 (T)$ below 70 K, we confirmed that the calculated data  agree with  both the experimental and our calculated results.
These results lead us to believe that the present analysis and the parameters obtained here can provide an appropriate characterization for  $K_1 (T)$ and $K_2 (T)$ at the quantitative level at least above 100 K. \par

\begin{figure}
\begin{center}
\includegraphics[height=6.0cm]{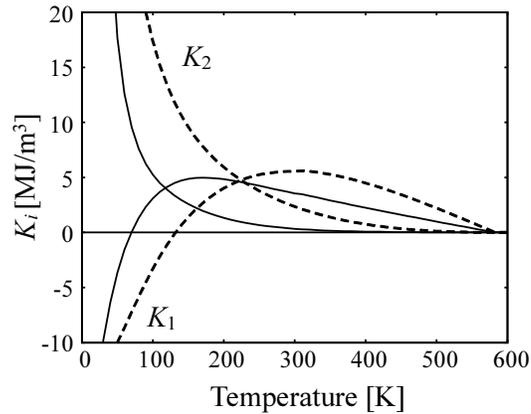}
\end{center}
\caption{Calculated anisotropy constants $K_1$ and $K_2$ (solid lines) as a function of temperature obtained  using the same parameters as in Fig. \ref{f1} (dashed lines) but with the half value of $H_\mathrm{ex}$.}
\label{f2}
\end{figure}

\begin{figure}
\begin{center}
\includegraphics[height=6.0cm]{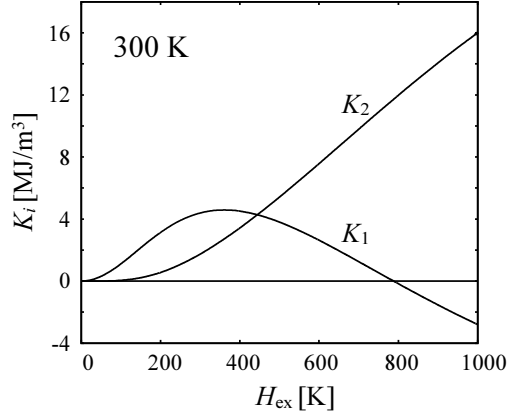}
\end{center}
\caption{$H_\mathrm{ex}$ dependence of the anisotropy constants $K_1$ and $K_2$ at $T= 300\ \mathrm{K}$.  The CEF parameters are the same as in Fig. \ref{f1}.}
\label{f3}
\end{figure}

Next we proceed to examine how  $K_1 (T)$ and $K_2 (T)$ are affected by the variation of amplitude of $H_\mathrm{ex} (T)$.
Figure \ref{f2} shows the calculated $K_1 (T)$ and $K_2 (T)$ using the half value of $H_\mathrm{ex}$ with the other parameters unchanged.
The results in Fig. \ref{f1} are also shown for comparison.  One can see a steep reduction in both $K_1 (T)$ and $K_2 (T)$ near a temperature of 70 K.
This in turn leads to the result that, for temperatures above 200 K, both  $K_1$ and $K_2$ values  become smaller than those in Fig. \ref{f1}.
To see the $H_\mathrm{ex}$ dependence of $K_1$ and $K_2$ values, we plot in Fig.
\ref{f3} these values as a function of $H_\mathrm{ex}$ at $T= 300$ K.
In these calculations, we neglect the Fe contribution $K_1^\mathrm{Fe} (T)$
to focus on only the effects of $H_\mathrm{ex}$ on the 4f-related anisotropy constants.
One finds that  $K_2$ exhibits a monotonic increase with $H_\mathrm{ex}$ whereas  $K_1$ has a peak at around $H_\mathrm{ex}=300\ \mathrm{K}$ and goes into the negative region above $H_\mathrm{ex} = 800\ \mathrm{K}$.
The negative value of $K_1$ for the high-$H_\mathrm{ex}$ region is naturally understood by considering the fact that  $K_1$ starts from a negative value at $T = 0$;
that is, the high limit of $H_\mathrm{ex}$ effectively corresponds to the low-temperature limit.
Further attention should be paid to the low-$H_\mathrm{ex}$ region below $H_\mathrm{ex}=300\ \mathrm{K}$,
where both  $K_1$ and $K_2$ values degrade with decreasing $H_\mathrm{ex}$.
This is because the 4f electron cloud approaches a spherical one, in accordance with the decrease of $H_\mathrm{ex}$, resulting in an insensitivity to the CEF and a decrease of the anisotropy energy.  From this behavior, one should recognize that the temperature dependence of the anisotropy constants $K_1$ and $K_2$ comes not only from the thermal fluctuation of the 4f moments but also from the strength of $H_\mathrm{ex} (T)$ through its temperature dependence as $\braket{\bm{S}_\mathrm{Fe}}$.
This can explain the vanishing of $K_1$,  $K_2$, and hence $\HC$ at the Curie temperature.
Further, even at  lower temperature,  the anisotropy constants may become small  around the grain boundaries where the exchange fields become small owing to the smaller coordination number or to some other factors such as stresses and lattice defects.
For these reasons, at around the operating temperature of electric vehicle motors (～500 K),
the decrease in the amplitude of $H_\mathrm{ex}$ in addition to the reduction of $\braket{\bm{S}_\mathrm{Fe}}$ owing to its temperature dependence may significantly influence magnet performance. \par

\begin{figure}
\begin{center}
\includegraphics[height=6.0cm]{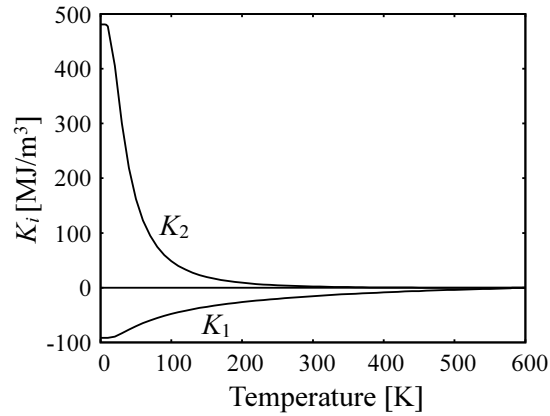}
\end{center}
\caption{Calculated anisotropy constants $K_1$ and $K_2$ as a function of temperature obtained  using $A_2^0=-820\ \mathrm{K/a_0{}^2}$, $A_4^0=A_6^0=0$, and $H_\mathrm{ex}=364\ \mathrm{K}$.}
\label{f4}
\end{figure}

Finally, we investigate the effects of variation of the CEF parameters on $K_1 (T)$ and $K_2 (T)$.  As pointed out in our previous work,\cite{15} there may be a possibility that  $A_2^0$ exhibits  a negative value at the grain surfaces, depending on whether their characteristics differ from those of the bulk.   Figure \ref{f4} shows the behaviors of $K_1 (T)$ and $K_2 (T)$  using the parameters $A_2^0=-820\ \mathrm{K/a_0{}^2}$,
%$A_4^0=0\ \mathrm{K/a_0{}^4}$,
%$A_6^0=0\ \mathrm{K/a_0{}^6}$,
$A_4^0=A_6^0=0$,
and $H_\mathrm{ex}=364\ \mathrm{K}$.
Clearly seen is that  $K_1 (T)$ exhibits a negative value for any temperature and that $K_2 (T)$ vanishes at  temperatures above 200 K,
which implies that the system favors planar anisotropy above 200 K.
Thus the prediction of $K<0$ when $A_2^0<0$ reported in the previous study is justified in the room-temperature region,
from which Mitsumata et al.\cite{17} suggested that  $\HC$ drastically degrades owing to the negative $K$ at the surface. \par

To conclude this paper, we summarize our study as follows.  To identify the possible mechanism of coercivity $(\HC)$ degradation of Nd-Fe-B sintered magnets, we have investigated how the magnetic anisotropy is affected by a variation of exchange fields and the CEF acting on the 4f electrons.  To this aim, the temperature-dependent anisotropy constants $K_1 (T)$ and $K_2 (T)$ were calculated based on  crystal field theory.  Using a certain set of CEF parameters and exchange field strength, we can obtain $K_1 (T)$ and $K_2 (T)$ curves that reproduce well the experimental data for bulk $\mathrm{Nd_2 Fe_{14} B}$.
It was found that, with decreasing exchange field strength, both values  decrease as a result of the lower asphericity of the 4f electron cloud.  This feature makes $K_1 (T)$, $K_2 (T),$ and hence $\HC$ vanish at the Curie temperature, where the exchange field strength tends to zero.
Even at a lower temperature,  the anisotropy constants may become small  around the grain boundaries where the exchange fields become small owing to the smaller coordination number.
For these reasons, at around the operating temperature of  electric vehicle motors (～500 K), the decrease in the amplitude of $H_\mathrm{ex}$ in addition to the reduction of $\braket{\bm{S}_\mathrm{Fe}}$ owing to its temperature dependence may significantly influence magnet performance.
It is also confirmed that the negative value of $A_2^0$, the leading CEF parameter, results in a planar anisotropy at room temperature.  Thus the prediction of $K<0$ when $A_2^0<0$ reported in the previous study is justified at the room-temperature region.

\acknowledgment
%If you need acknowledgement(s), use the \verb|\acknowledgement| command. We have prepared variants of this command as \verb|\acknowledgements|, \verb|\acknowledgment|, and \verb|\acknowledgments|.
This work was supported by JST--CREST.


\begin{thebibliography}{99}
%\bibitem{apex} The abbreviation for APEX must be ``Appl. Phys. Express" in the reference list.
\bibitem{1} M. Sagawa, S. Hirosawa, K. Tokuhara, H. Yamamoto, S. Fujimura, Y. Tsubokawa, and R. Shimizu, J. Appl. Phys. \textbf{61}, 3559 (1987).
\bibitem{2} M. Sagawa, S. Hirosawa, H. Yamamoto, S. Fujimura, and Y. Matsuura, Jpn. J. Appl. Phys., Part \textbf{1 26}, 785 (1987).
\bibitem{3} J. F. Herbst, Rev. Mod. Phys. \textbf{63}, 819 (1991).
\bibitem{4} Fidler and K. G. Knoch, J. Magn. Magn. Mater. \textbf{80}, 48 (1989).
\bibitem{5} F. Vial, F. Joly, E. Nevalainen, M. Sagawa, K. Hiraga, and K. T. Park, J. Magn. Magn. Mater. \textbf{242--245}, 1329 (2002).
\bibitem{6} W. F. Li, T. Ohkubo, and K. Hono, Acta Mater. \textbf{57}, 1337 (2009).
\bibitem{7} W. Mo, L. Zhang, Q. Liu, A. Shan, J. Wu, and M. Komuro, Scr. Mater. \textbf{59}, 179 (2008).
\bibitem{8} Y. Shinba, T. J. Konno, K. Ishikawa, and K. Hiraga, J. Appl. Phys. \textbf{97}, 053504 (2005).
\bibitem{9} T. G. Woodcock, Y. Zhang, G. Hrkac, G. Cuta, N. M. Dempsey, T. Schrefl, O. Gutfleisch, and D. Givord, Scripta Mater. \textbf{67}, 536 (2012).
\bibitem{10} K. Hono and H. Sepehri-Amin. Scr. Mater. \textbf{67}, 530 (2012).
\bibitem{11} Sepehri-Amin, T. Ohkubo, T. Shima, and K. Hono, Acta Mater. \textbf{60}, 819 (2012).
\bibitem{12} H. Kronm$\ddot{\rm u}$ller, K.-D. Durst, and G. Martinek, J. Magn. Magn. Mater. \textbf{69}, 149 (1987).
\bibitem{13} A. Sakuma, S. Tanigawa, and M. Tokunaga, J. Magn. Magn. Mater. \textbf{84}, 52 (1990).
\bibitem{14} R. Fischer. T. Schrefl, H. Kronm$\ddot{\rm u}$ller,  and J. Fidler, J. Magn. Magn. Mater. \textbf{153}, 35 (1987).
\bibitem{15} H. Moriya, H. Tsuchiura, and A. Sakuma, J. Appl. Phys., \textbf{105}, 07A740 (2009).
\bibitem{16} G. Hrkac, T. G. Woodcock, C. Freeman, A. Goncharov, J. Dean, T. Schrefl, and O. Gutfleisch, Appl. Phys. Lett. \textbf{97}, 232511 (2010).
\bibitem{17} C. Mitsumata, H. Tsuchiura, and A. Sakuma, Appl. Phys. Express \textbf{4}, 113002 (2011).
\bibitem{18} M. Yamada, H. Kato, H. Yamamoto, and Y. Nakagawa, Phys. Rev. B \textbf{38}, 620 (1988).
\bibitem{19} S. Hirosawa, Y. Matsuura, H. Yamamoto, S. Fujimura, and M. Sagawa, J. Appl. Phys. \textbf{59}, 873 (1986).
\bibitem{20} For the CEF parameters and $H_\mathrm{ex}$, we make use of $\mathrm{K}$ (Kelvin) as  a unit of energy, following the manner of Ref. 18.
\bibitem{21} O. Yamada, H. Tokuhara, F. Ono, M. Sagawa, and Y. Matsuura, J. Magn. Magn. Mater., \textbf{54--57}, 585 (1986).
\end{thebibliography}
\end{document}